\documentclass[a4paper,fleqn,usenatbib]{mnras}

\usepackage{mathtools}
\usepackage{listings}
\usepackage{color} 
\usepackage{listings}
\usepackage{hyperref}
\usepackage{newtxtext,newtxmath}
\usepackage[T1]{fontenc}
\usepackage{ae,aecompl}
\usepackage{threeparttable} 
\usepackage{deluxetable} 
\usepackage{graphicx}	
\usepackage{amsmath}	
\usepackage{amssymb}	
\newcounter{question}
\definecolor{dkgreen}{rgb}{0,0.6,0}
\definecolor{gray}{rgb}{0.5,0.5,0.5}
\definecolor{mauve}{rgb}{0.58,0,0.82}

\usepackage{natbib,multicol,rotating}

\newcommand{\gs}{\mathrel{\raise0.35ex\hbox{$\scriptstyle >$}\kern-0.6em
\lower0.40ex\hbox{{$\scriptstyle \sim$}}}}
\newcommand{\ls}{\mathrel{\raise0.35ex\hbox{$\scriptstyle <$}\kern-0.6em
\lower0.40ex\hbox{{$\scriptstyle \sim$}}}}
\def\beq{\begin{equation}\begin{aligned}}
\def\eeq{\end{aligned}\end{equation}}

\graphicspath{{plots/}}

\title[Oscillating Artifacts]{Windowing Artifacts Likely Account for Recent Claimed Detection of Oscillating Cosmic Scale Factor}
\author[Brownsberger, Stubbs, Scolnic ]{Sasha R. Brownsberger $^{1}$  \thanks{sashabrownsberger@g.harvard.edu}, Christopher W. Stubbs$^{1,2}$, Daniel M. Scolnic$^{3}$ \\ 
$^1$ Department of Physics Harvard University, Cambridge, MA 02138 , USA \\ 
$^2$ Harvard-Smithsonian Center for Astrophysics, Cambridge, MA 02138, USA \\ 
$^3$ Duke University, Department of Physics, Durham, NC 27708, USA}

\date{Accepted XXX. Received YYY; in original form ZZZ}

\pubyear{2020}

\begin{document}
\label{firstpage}
\pagerange{\pageref{firstpage}--\pageref{lastpage}}
\maketitle

\begin{abstract}
Using the Pantheon data set of Type Ia supernovae, \cite{Ringermacher2020} (R20 henceforth) report a $~2\sigma$ detection of oscillations in the expansion history of the universe.  
Applying the R20 methodology to simulated Pantheon data, we determine that these oscillations likely arise from analysis artifacts.  The uneven spacing of Type Ia supernovae in redshift space and the complicated analysis method of R20 impose a structured throughput function.  
When analyzed with the R20 prescription, about $11\%$ of artificial $\Lambda$CDM data sets produce a stronger oscillatory signal than the actual Pantheon data. 
The study conducted by R20 is a wholly worthwhile endeavor.  However, we believe that the detected oscillations are not due to an oscillating cosmic scale factor and are instead artifacts of the data processing. Our results underscore the importance of understanding the false `signals' that can be introduced by complicated data analyses.  
\end{abstract}

\section{Introduction and Background}
Since the initial discovery of dark energy (DE) by \cite{Riess1998, Perlmutter1999}, observations of Type Ia supernovae (SNe Ia) have been integral in establishing the canonical $\Lambda$CDM cosmological model.  
In the $\Lambda$CDM model, the present energy density of our flat universe is dominated by cosmologically constant DE ($\Lambda$) and non-relativistic, collisionless (`cold') dark matter (CDM).  
Though some tensions between predictions and observations exist \citep{Weinberg2015, Verde2019}, this simple model has successfully predicted many cosmological and astrophysical signals \citep{Peter2012, Mortonson2013}.  

However, despite the success of the $\Lambda$CDM model, the physical identifies of $\Lambda$ and CDM remain unsettled.  
Researchers continue to search for deviations from the predictions of $\Lambda$CDM in many data sets, including an ever-increasing archive of SNe Ia.   
In particular, some non-canonical cosmological models, such as those discussed by \cite{Barenboim2005, Xia2005, Feng2006, Lazkoz2010, Wang2017}, predict that the true expansion of the universe might oscillate around the predictions of $\Lambda$CDM.  
Respectively using the Gold \citep{Reiss2007}, Union \citep{Kowalski2008}, Constitution \citep{Hicken2009} and Pantheon \citep{Scolnic2018} data sets of SNe Ia, \cite{Jain2007}, \cite{Liu2009}, \cite{Lazkoz2010}, and \cite{Brownsberger2019} search for evidence of such oscillations in cosmic expansion.  Though they utilize a diversity of data sets and statistical methods, those analyses universally report no evidence of oscillations in the rate of cosmic expansion. 

Contrary to those previous findings, \cite{Ringermacher2015} and \cite{Ringermacher2020} (R15 and R20 henceforth) claim to identify damped oscillations in the universe's recent expansion history.
Combining data of radio galaxies and SNe Ia \citep{Conley2011, Daly2004, Reiss2004} into a `CDR' data set, R15 claim to detect cosmic oscillations in the universe's scale factor.  
Using the Pantheon data set of type Ia supernovae, R20 build on R15 to claim a detection of an oscillating scale factor with a total statistical significance of at least $2\sigma$.

Such a detection of oscillatory cosmic expansion would mark an enormous paradigm shift in our understanding of the physics of the universe, changing the canonical model that has held since the first identification of DE.
The work of R20 is entirely worthwhile.  Their results should be seriously considered and appropriately scrutinized.



Replicating the analysis method of R20 and applying it to simulated data, we find that there is an $11\%$ chance that the Pantheon data observed in a $\Lambda$CDM universe would produce a stronger oscillatory signal than that which R20 detect.  
Our measurement does not include a statistical `trials factor' penalization for the various tunable parameters in the R20 analysis, and the significance of the detected oscillations is therefore less than our reported metric.  
The oscillations noted by R20 are likely data analysis artifacts - the signature of a throughput function that consists of the uneven spacing of the Pantheon SNe in redshift and their sequencing of filtering and differentiation analysis steps.


In Section \ref{sec:replicate} below, we describe our replication of the R20 analysis.  In Section \ref{sec:random}, we describe our generation of the artificial data and the assessment of the consistency of the real data with $\Lambda$CDM.  We detail our conclusions in Section \ref{sec:conclusion}. 

\section{Replicating the R20 Results} \label{sec:replicate} 
In this Section, we describe our replication of the R20 analysis and the R20 results.

\subsection{Inferring Cosmic Time and Residual Scale Factor Derivative for SNe Ia}  \label{sec:processing}
R20 search for oscillations by transforming the standard Hubble diagram (brightness vs. redshift) into plots of scale factor vs time.  They claim that such a plot enables a model-independent study of the universe's expansion history. 

The Pantheon data set of Type Ia supernovae consists of measured redshifts, $z_i$, distance moduli, $\mu_i$, and distance modulus uncertainties, $\sigma_{\mu, i}$.  The subscripts identify each SNe in order of increasing $z_i$.  These are the directly measured quantities from which cosmologists infer cosmic expansion.

From these measured quantities, R20 note oscillations in non-standard inferred quantities: the normalized cosmic time since the end of inflation and the residual in the time derivative of the scale factor.
They infer measurements of cosmic time, which we denote $t_i$, by approximating an integral in luminosity distances with a discrete sum.  
They approximate residual time derivatives in scale factor by integrating in luminosity distance to determine $t_i$, subtracting the predictions of $\Lambda$CDM to measure residuals, binning these residuals, taking a discrete derivative of these binned values, smoothing these binned derivatives with distinct smoothing kernels, and subtracting the two smoothings.

We make this series of operations explicit in our notation of the inferred residual  time derivative scale factor, which we denote $\Delta G(d \overline{\Delta a_i}/dt)$.  Here, the inner $\Delta$ indicates a residual, $\overline{(\cdot)}$ indicates binning, $d(\cdot)/dt$ indicates discrete time differentiation, and $\Delta G$ indicates the differences between two Gaussian smoothings. 

Throughout the rest of this Section, we describe how we inferred $t_i$ and $\Delta G(d \overline{\Delta a_i}/dt)$ from $z_i$ and $\mu_i$. 
We measured the cosmological scale factors, $a_i$, and luminosity distances, $d_{L, i}$, using the standard relations: 
\begin{equation} 
a_i = \frac{1}{1+ z_i}  , 
\end{equation}  
and 
\begin{equation} 
d_{L, i}= 10 ^ {(\mu_i - 25) / 5} \textrm{ Mpc}. 
\end{equation}  
The scaled luminosity distances, $Y_i$, were defined by
\begin{equation} 
Y_i = a_i \frac{d_{L, i}}{D_H} ,
\end{equation}  
where $D_H  = c / H_0$, $c$ is the speed of light, and $H_0$ is the Hubble constant.  
We determined the scaled $Y$ separations between measured SNe, $\Delta Y_i$, via the relation: 
\begin{equation}
a_i \Delta Y_i = a_i (Y_i - Y_{i-1}) . 
\end{equation}  
We calculated the normalized cosmological times, $t_{i, raw}$, by approximating an integral over cosmic time via a discrete sum: 
\begin{equation} 
t_{i, raw} = 1 - \int_0^{z_i} a(t) dY \simeq 1 - \sum_{j=1}^{i} a_j \Delta Y_j . 
\end{equation} 
We corrected the raw cosmological times using the relation: 
\begin{equation} 
t_{i} = \alpha_{\textrm{Pan to CDR}} (t_{i, raw}- t_{corr}) , 
\end{equation}
where $t_i$ are the corrected cosmological times.  
According to R20, $t_{corr}$ corrects for the fact that the first measurement of $t$ is at the first SN where $a$ is not equal to its present day value, and $\alpha_{\textrm{Pan to CDR}}$ is a scaling to match the Pantheon $t$ range to the $t$ range of the CDR data set studied in R15.  Following R20, we used $t_{corr} = 0.009579$ and $\alpha_{\textrm{Pan to CDR}}=1.041$.  

We calculated the residual scale factors, $\Delta a_i$, by subtracting from the measured $a_i$ values the canonical values of $a_i$ determined from $t_{i}$: 
\begin{equation}
\Delta a_i = a_i - a_{\Lambda CDM} (t_{i}) . 
\end{equation} 
We defined the canonical scale factor, $a_{\Lambda CDM}$, for a given cosmic time, $t$, by the integral relation
\begin{equation} \label{eq:aCanon} 
t = 1 - \int_0^{1 / a_{\Lambda CDM} - 1} dz' \frac{1}{(1+z' )\sqrt{\Omega_M (1+ z') ^ 3 + \Omega_{\Lambda} }} . 
\end{equation} 
Copying R20, we set $\Omega_M  = 0.27$ and $\Omega_{\Lambda} = 0.73$.  We calculated $a_{\Lambda CDM} (t_{i})$ for each $t_{i}$ by interpolating over an array of $t$ values calculated at $N_{a, interp} = 1001$ $a_{\Lambda CDM}$ values evenly distributed over the physically relevant range of $a_{\Lambda CDM} \in [0, 1]$.  With $N_{a, interp} = 1001$, our interpolated values of $a_{\Lambda CDM} (t_{i})$ converged to within $0.01\%$ of their true values. 



To bin the inferred scale factor residuals, we divided the $t$ space ($0$ to $1$) into $N_{bin} = 128$ bins of equal size and calculated the mean $\Delta a_i$ value in each $t$ bin, $\overline{\Delta a_i}$ .  We computed the wide baseline derivative of $\overline{\Delta a_i}$, $d\overline{\Delta a_i}/dt$, following Equation (1) of R20: 
\begin{equation}
\frac{d{\overline{\Delta a_i}}}{dt} = \frac{\overline{\Delta a_{i+n/2}} - \overline{\Delta a_{i-n/2}}}{n \Delta t} . 
\end{equation} 
As in R20, $n = 8$ and $\Delta t = 1/128$.  For the first [last] $n/2$ bins, the lower [upper] $\Delta a_{i}$ value was set to the first [last] bin and the $n$ in the denominator was set equal to the number of bins over which the derivative was measured.   

We smoothed the $d\overline{\Delta a_i}/dt$ values using a Gaussian kernel.  We denote these smoothed derivatives as $G_k(d \overline{\Delta a_i}/dt)$ where the $k$ index denotes the width of the Gaussian kernel in $t$: 
\begin{equation} \label{eq:smoothDef}
 G_k(x) =\frac{ \sum_{j=0}^{N_{bin}} x\ G( \frac{t_i - t_j}{k } ) }{ \sum_{j=0}^{N_{bin}} G( \frac{t_i - t_j}{k } )} , 
\end{equation} 
where
\begin{equation} \label{eq:GDef}
G (a) = \frac{1}{\sqrt{2 \pi} \times 0.37} e^{(- a^2 / (2 \times 0.37 ^ 2))} \ .
\end{equation} 
We based Equations \ref{eq:smoothDef} and \ref{eq:GDef} on the definition of the \lstinline{ksmooth} function of the \lstinline{Mathcad} software, as that is the smoothing function used by R20.  We believe the value of 0.37 is an approximation of one e-folding, $1/e$.  

\begin{figure}
\centering 
\includegraphics[width=0.8\columnwidth]{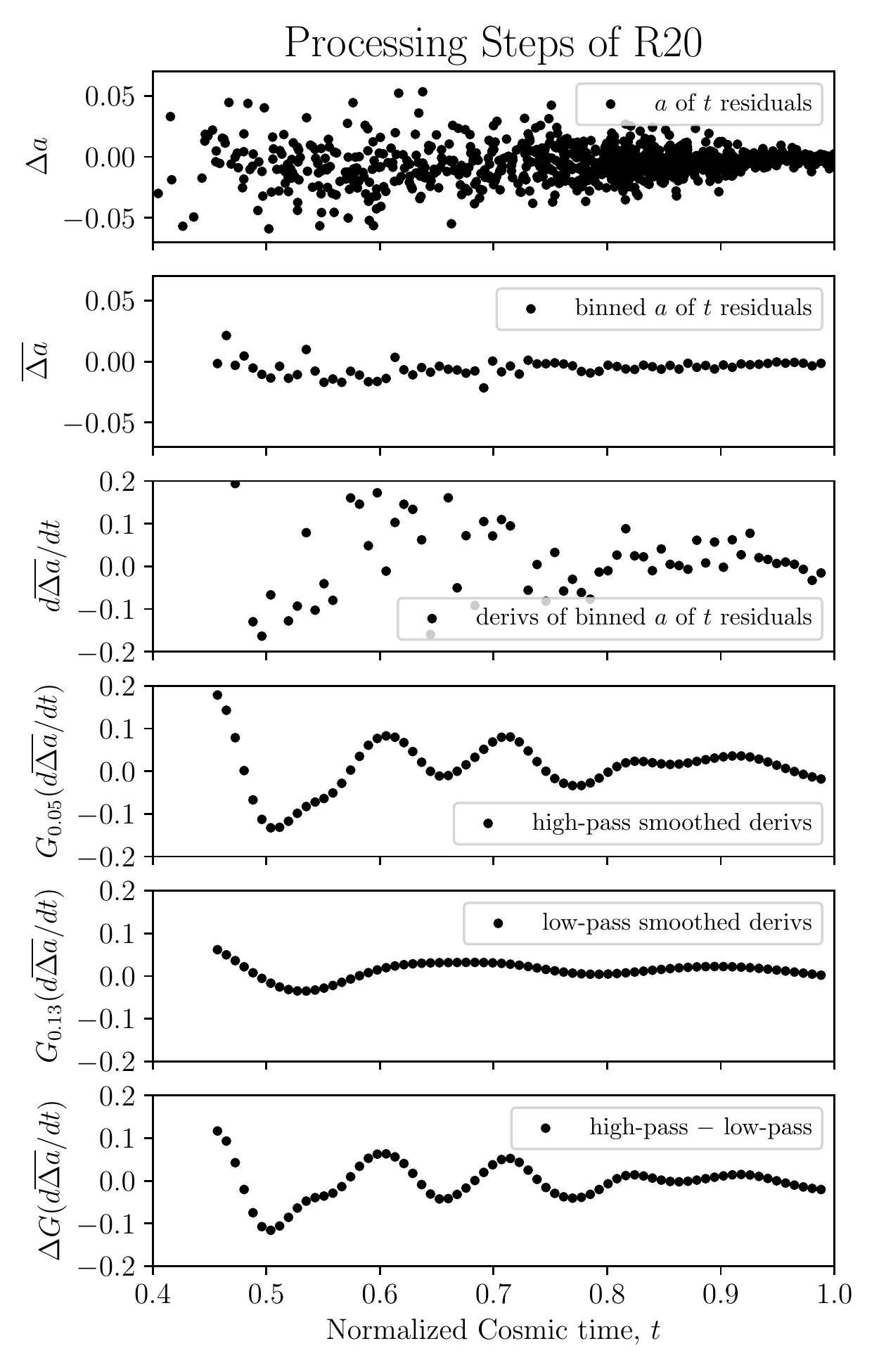} 
\caption{The results of applying the analysis of R20 to the Pantheon data set of Type Ia supernovae \citep{Scolnic2018}.  Starting at the top with the inferred scale factor, $\Delta a$ vs inferred cosmic time, $t_i$, each plot shows the data after one additional step of the analysis discussed in Section \ref{sec:processing}.  The bottom plot, showing $\Delta G(d \overline{\Delta a_i}/dt)$ vs $t_i$, resembles the results in Figure 2 of R20 and displays oscillations.} \label{fig:trueDataPlot}
\end{figure}

Our final result, $\Delta G(d \overline{\Delta a_i}/dt)$, is the difference between $G_k(d \overline{\Delta a_i}/dt)$ with two kernels: 
\begin{equation} 
\Delta G(\frac{d \overline{\Delta a_i}}{dt}) = G_{k1}(\frac{d \overline{\Delta a_i}}{dt}) - G_{k2}(\frac{d \overline{\Delta a_i}}{dt}) \ .
\end{equation} 
Here, as in R20, we set $k1 = 0.05$ and $k2 = 0.13$.  
We emphasize that $\Delta G(d \overline{\Delta a_i}/dt)$ is not a true measurement of the residual of the time derivative of the scale factor, $\Delta \dot{a}$.  Rather, $\Delta G(d \overline{\Delta a_i}/dt)$ represents an attempt to infer $\Delta \dot{a}$ through a series of data processing steps.  The relation between $\Delta G(d \overline{\Delta a_i}/dt)$ and $t_i$ describes the cosmic relation between $\Delta \dot{a}$ and $t$ viewed through a structured windowing function.   

\subsection{Results from the True Pantheon Data}

In Figure \ref{fig:trueDataPlot}, we show the intermediate results of the data processing steps described in Section \ref{sec:processing}.  To best replicate Figure 2 of R20, we measured only those bins with $t \geq 0.46 $ and $t \leq 1 $.  The bottom panel, showing our calculation of $\Delta G(d \overline{\Delta a_i}/dt)$ vs $t_{i}$, represents our best replication of Figure 2 in R20.  

We find a damped oscillatory relation between $\Delta G(d \overline{\Delta a_i}/dt)$ and $t_{i}$.  
The oscillation amplitude decreases in $t$, with the largest excursion of $|\Delta G(d \overline{\Delta a_i}/dt)| \simeq 0.12$ occurring around $t \simeq 0.5$.  
We found that such oscillations around in $\Delta \dot{a}$ in $t$ would correspond to similar oscillations in distance modulus residual, $\Delta \mu$, in redshift, $z$.  We estimate that, for oscillations of the magnitude shown in Figure \ref{fig:trueDataPlot}, there would be an oscillatory signal in $\Delta \mu$ vs $z$ with a peak amplitude of about 10 millimags.  Comparing this prediction to the constraints shown in Figure 6 of \cite{Brownsberger2019}, the Pantheon data are unable to rule-out such a small oscillatory signal in $\Delta \mu$ vs $z$.  

Having demonstrated the consistency of our analysis with that of R20, we next show how similar oscillatory signals can arise from the above analysis applied to Pantheon-like artificial data obtained in a canonical $\Lambda$CDM universe.

\section{Demonstrating how the Claimed Oscillatory Signal is Generic, and not an Indication of  Cosmic Oscillations} \label{sec:random} 
The final data shown in the lower plot of Figure \ref{fig:trueDataPlot} are \emph{not} direct measurements of the residual time derivative of the scale factor $\Delta \dot{a}$.  
Rather, they represent an inference of $\Delta \dot{a}$  acquired through a serious of operations.  
Those operations conspire with the sampling of observed SNe in redshift to produce a complicated windowing function which itself carries structure. 

In this section, we describe how we produced simulated Pantheon-like data sets and demonstrate that random data realizations distributed around the $\Lambda$CDM cosmology can generate the same sorts of oscillations identified in the bottom plot of Figure \ref{fig:trueDataPlot}. 

\subsection{Generating Randomized Pantheon-Like Data} 
 We applied the same procedure discussed in Section \ref{sec:replicate} to randomized versions of the Pantheon data set.  
For the $i^{th}$ Pantheon SN, we determined a randomized $\mu_i$ by drawing from a normal distribution with mean equal to the background $\Lambda$CDM predicted $\mu_i$ and standard deviation equal to the reported $\sigma_{\mu,i}$.  The $z_i$ of each SN is well determined, and was left unchanged in the randomization. This preserves the window function in redshift.  Each randomization produced a new set of 1048 $\mu_i$ values at the same $z_i$ positions.  

\begin{figure}
\centering 
\includegraphics[width=1.0\columnwidth, angle = 0]{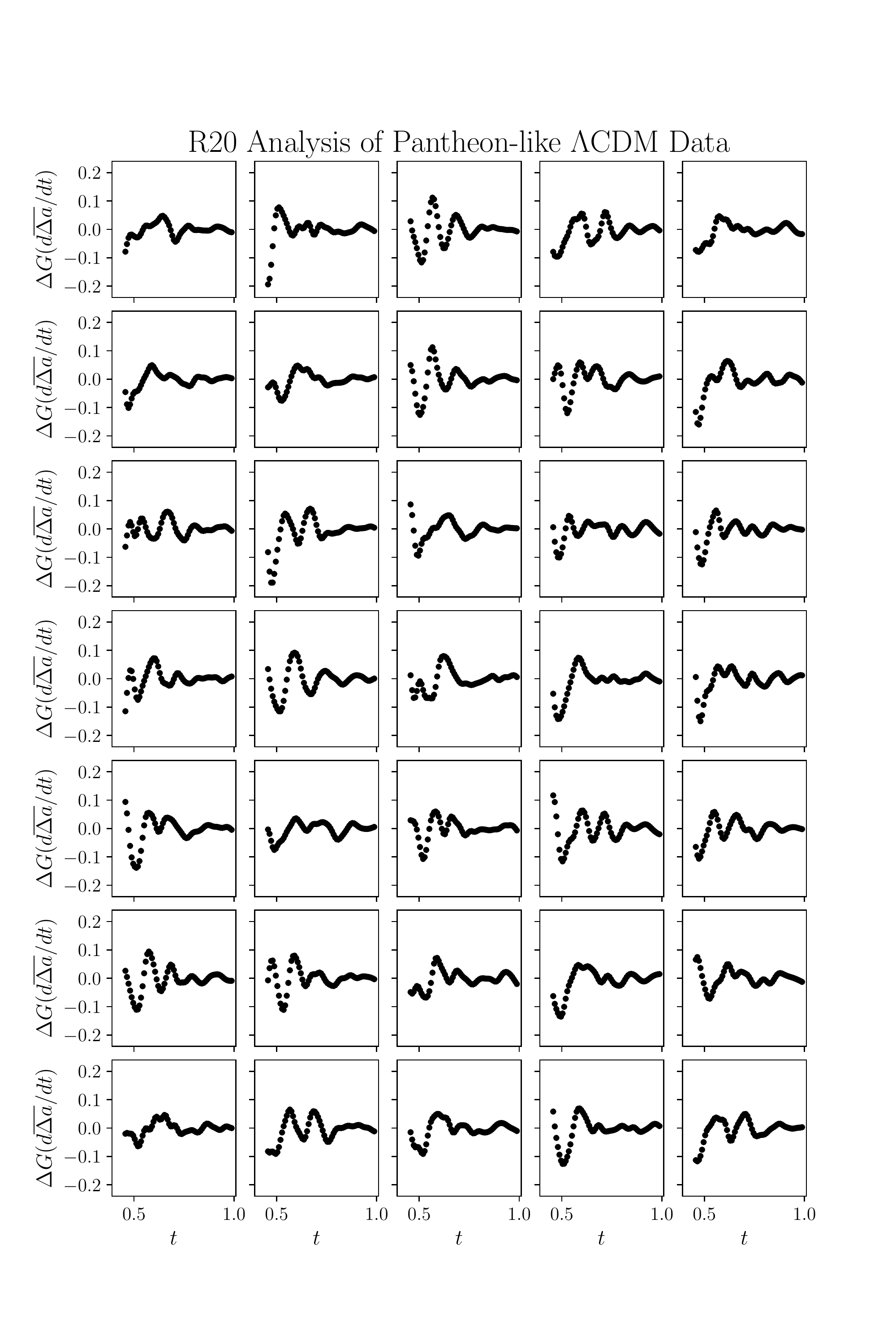} 
\caption{Thirty-four plots of artificial Pantheon-like data sets and the plot of the real Pantheon data all processed using the analysis in Section \ref{sec:processing}.  We identify which panel displays the true data at the end of this caption.  
Because we randomized the distance moduli of these Pantheon-like data sets around $\Lambda$CDM, they are definitionally bereft of non-canonical cosmic structure.  The oscillations exhibited in all but one of the above plots are nothing more than artifacts of the data processing.  Because the oscillations of the true Pantheon data are not clearly distinct from the oscillations in the artificial Pantheon-like data, we argue that the oscillations identified by R20 could reasonably result from the Pantheon data observed in a canonical $\Lambda$CDM cosmology.  
In this figure, the plot of the true Pantheon data is displayed in the fifth row of the fourth column.} \label{fig:randCanonPlot}
\end{figure}

Any fundamental cosmic oscillation buried in the true Pantheon data does not exist in these artificial Pantheon-like data sets.  
By pushing each randomized data set through the same processing steps described in Section \ref{sec:replicate}, we generated a plot of $\Delta G(d \overline{\Delta a_i}/dt)$ vs $t_{i}$ for Pantheon-like data from which any non-$\Lambda$CDM cosmic structure (oscillatory or otherwise) has been removed.  We repeated this randomization $N_{R} = 10^4$ times.  

In Figure \ref{fig:randCanonPlot}, we show a representative subset of the $\Delta G(d \overline{\Delta a_i}/dt)$ vs $t_{i}$ plots of the randomized Pantheon-like data sets.  Many such plots (those shown and not shown in Figure \ref{fig:randCanonPlot}) display oscillations similar in size and wavelength to the oscillations observed in Figure \ref{fig:trueDataPlot}.  To underscore this point, we include the $\Delta G(d \overline{\Delta a_i}/dt)$ vs $t_{i}$ plot of the true Pantheon data amongst the plots of randomized Pantheon data in Figure \ref{fig:randCanonPlot}.  
As an informal test of the real data's consistency with random fluctuations around $\Lambda$CDM, one can identify which of the plots in Figure \ref{fig:randCanonPlot} appears to show the strongest oscillatory signal and check if that plot depicts real or artificial data.  We identify the real Pantheon data in the Figure's caption.

\subsection{The Frequency Spectra of Real and Artificial Pantheon Data}
\begin{figure}
\centering 
\includegraphics[width=0.9\columnwidth, angle = 0]{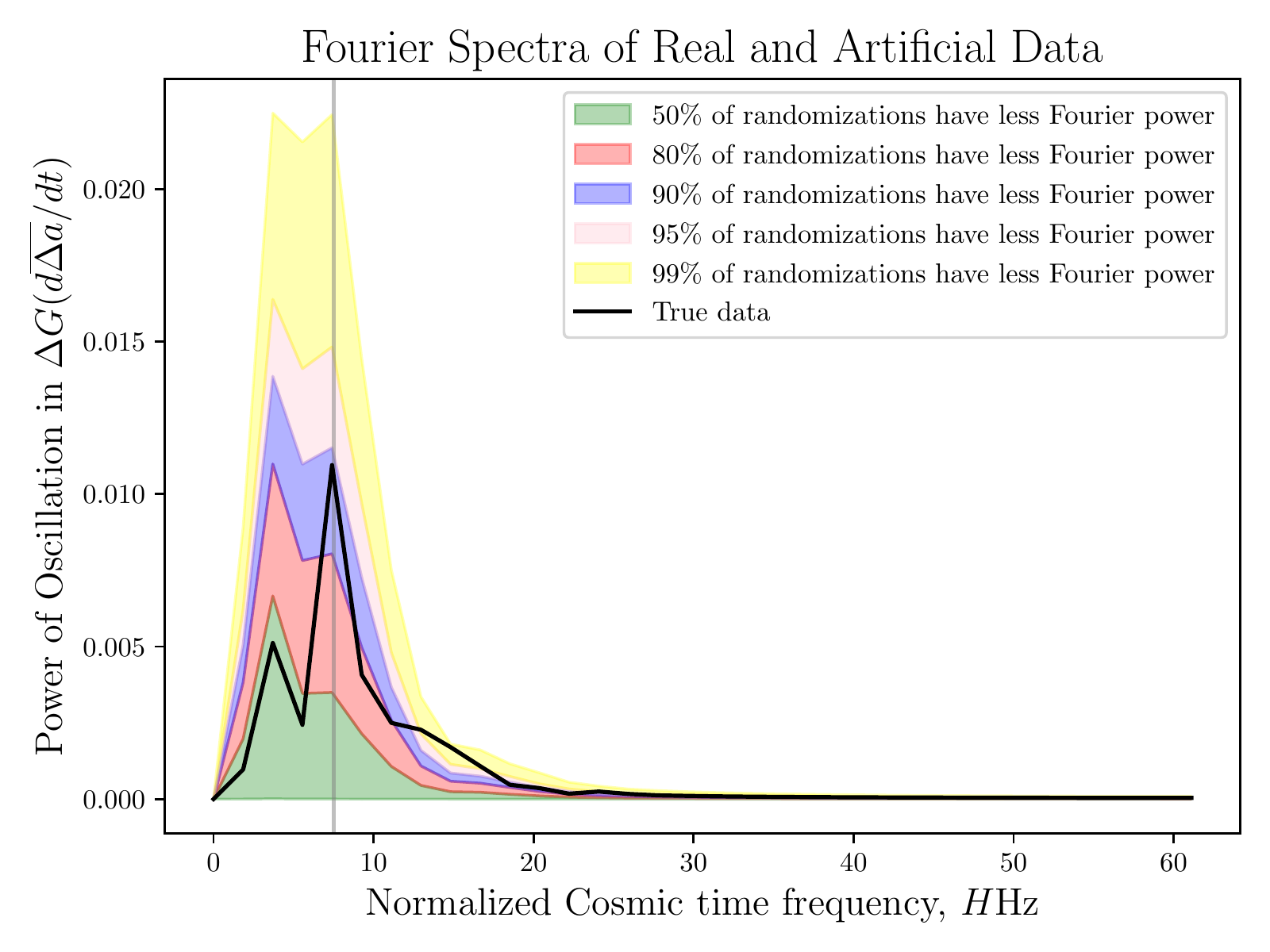} 
\caption{The power spectrum of the true Pantheon data (black line) and the distribution of power spectra of the randomized Pantheon-like data.  The $N_{R}=10^4$ randomized Pantheon-like data sets produce, at every frequency, a distribution of $N_{R}$ measurements of the power spectra that could result from random deviations around the $\Lambda$CDM cosmology.  At each frequency, the noted percentage of randomizations lie below the labeled contour.   For example, at each frequency, $50\%$ of randomizations have power below the green contour and $90\%$ of randomizations have power below the blue contour.} \label{fig:fourier}
\end{figure}

Examining the $\Delta G(d \overline{\Delta a_i}/dt)$ vs $t_{i}$ plots in Figure \ref{fig:randCanonPlot}, we cannot confidently distinguish the plot of the true Pantheon data from those of randomized data plots.  The oscillations in the real data appear consistent with apparent oscillations that could result from random fluctuations around the canonical $\Lambda$CDM cosmology.  

\begin{figure}
\centering 
\includegraphics[width=0.9\columnwidth, angle = 0]{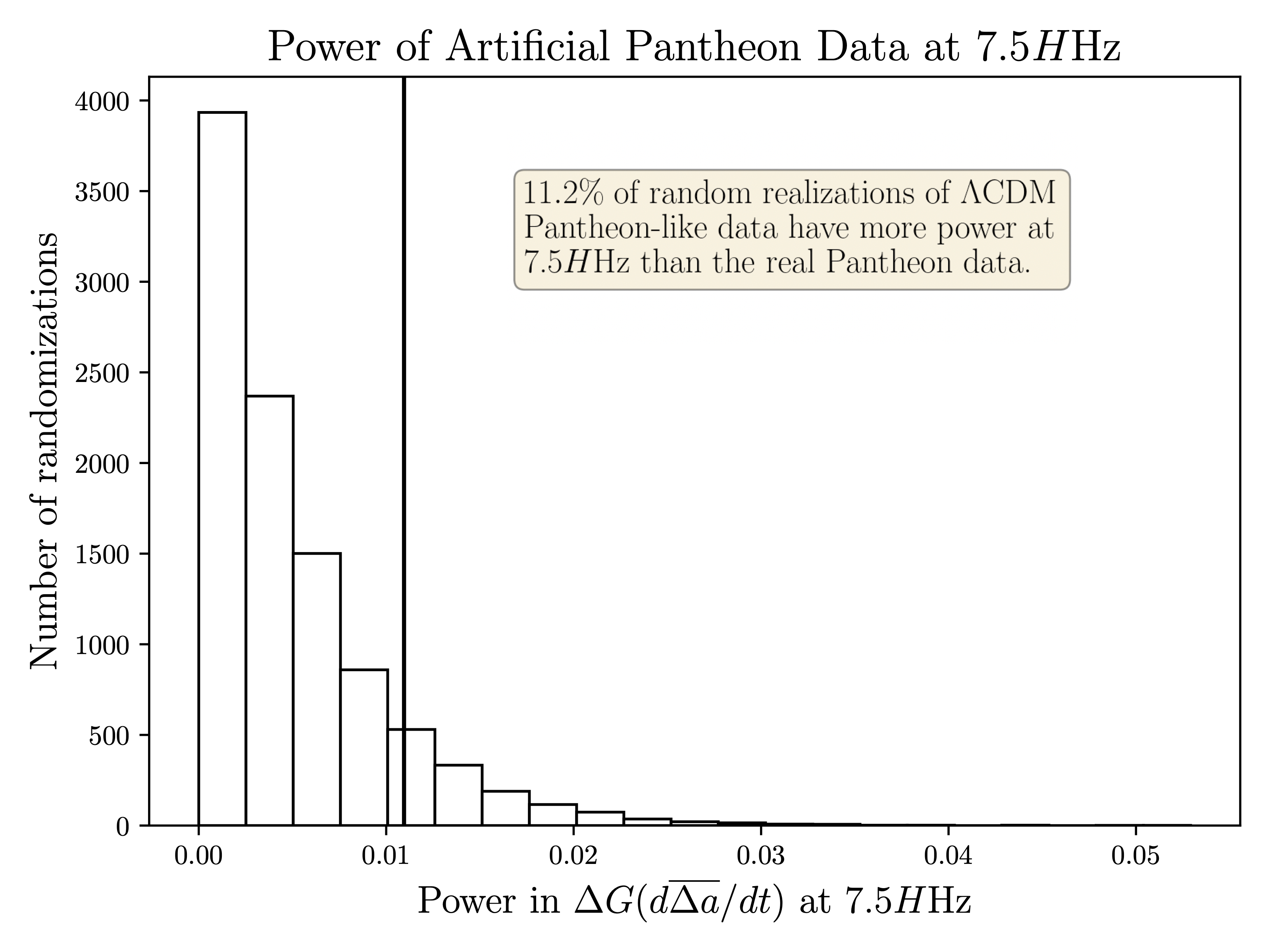} 
\caption{The distribution of Fourier power at $f = 7.5 H$Hz measured in the artificial Pantheon-like data randomly distributed around $\Lambda$CDM and analyzed according to the analysis of Section \ref{sec:replicate}.  
The power of the real Pantheon at this frequency subject to the same analysis is shown with the vertical black line. 
About $11\%$ of random data realizations have more Fourier power at this chosen frequency than the real Pantheon data.  
The Pantheon data, analyzed according to the methodology described by R20, shows no statistically significant evidence of oscillations in the rate of cosmic expansion.} \label{fig:hist}
\end{figure}

To make this qualitative observation quantitative, we computed power spectra of $\Delta G(d \overline{\Delta a_i}/dt)$ in $t_{i}$.  The data were binned in $t_{i}$ bins of equal size, and the data to be Fourier transformed was thus evenly spaced in time.  
We measured the Power Spectrum, $P_f$, of $\Delta G(d \overline{\Delta a_i}/dt)$ in $t_{i}$ via a standard discrete Fourier transform: 
\begin{equation} \label{eq:fourier}
P_f = \frac{1}{N_{bin}} \Big | \sum_{j=0}^{N_{bin}-1} \dot{\overline{\Delta a_j}}^{\Delta G} e^{-i 2 \pi / f  \ j / N_{bin}} \Big |^2 \ .
\end{equation} 
To avoid aliased modes, we measured the Fourier power in frequencies, $f$, from $0 H$Hz to $N_{bin} /4 = 32 H$Hz.  Replicating R20, 1 $H$Hz (`one Hubble Hertz') $ = 0.1023 h_{100} Gyr^{-1}$. 

Using Equation \ref{eq:fourier}, we computed the Power Spectrum of $\Delta G(d \overline{\Delta a_i}/dt)$ in $t_{i}$ for the true Pantheon data set and for the $N_{R}$ artificial Pantheon-like data sets.  We show the true Pantheon power spectrum and the distribution of artificial Pantheon-like power spectra in Figure \ref{fig:fourier}.  At its peak, the power spectrum of the true Pantheon data (black line) lies below the $90\%$ contour of the artificial power spectra (blue shading).  

R20 focus primarily on the frequency peak at $f = 7.5 H$Hz.  In Figure \ref{fig:hist}, we display a histogram showing the Fourier power at $f = 7.5H$Hz of the $N_{R}$ Pantheon-like artificial data sets and show where the power of the real Pantheon data lies in this histogram (black line).  
About $11\%$ of randomized Pantheon-like $\Lambda$CDM data sets have more power at the chosen frequency than the real Pantheon data. 


\section{Conclusions}\label{sec:conclusion} 
 We replicated the analysis of the Pantheon data set of SNe Ia described by R20 and we found a similar result: the inferred $\dot{a}$ residuals oscillate in the inferred cosmic time.  We show these results in Figure \ref{fig:trueDataPlot}.  
 
 We repeated this analysis on artificial Pantheon-like data sets with unchanged redshifts and with distance moduli randomly drawn from normal distributions centered at the canonical $\Lambda$CDM cosmology.  By definition, this randomization erased any cosmic oscillation signature that exists in the true Pantheon data.  Many plots of the R20 analysis applied to these randomized distributions (see Figure \ref{fig:randCanonPlot} for a representative subsample) display oscillations similar in amplitude and frequency to those identified in the real Pantheon data.  
 
To make this qualitative observation quantitative, we measured the power spectra of the real Pantheon data and of the artificial Pantheon-like data sets.  We showed these power spectra in Figure \ref{fig:fourier}.  R20 focus on the power spectrum peak at $f = 7.5 H$Hz.  In Figure \ref{fig:hist}, we showed the distribution of the artificial data sets' powers at this chosen frequency and where the true Pantheon data lies in this histogram.  
About $11\%$ of the Pantheon-like $\Lambda$CDM data sets have more power at this chosen frequency than the real Pantheon data when analyzed according to the prescription of R20. 
Our analysis used the same choice of tuned analysis parameters that R20 report, including the widths of the smoothed Gaussian kernels, the chosen frequency, and the number of time bins.  A robust measurement of the statistical significance of this $\simeq 11\%$ effect would also include a statistical penalization for these adjustable analysis parameters.  

There are potential sources of systematic error that neither we nor R20 consider.  Particularly, the Pantheon data set is a combination of distinct supernova survey projects, each of which carries its own imperfectly characterized systematic errors.  These inter-survey systematics inherit each individual survey's uneven distributions in redshift and on the sky.  
If the oscillations noted by R20 appeared to be more than data analysis artifacts, we would analyze the signal's robustness against these inter-survey systematics. 


 There is at least a one-in-ten chance that statistical fluctuations around the canonical $\Lambda$CDM cosmology would conspire with the windowing function of the R20 data analysis to produce a larger oscillatory signal than that which R20 report.  The apparent oscillatory signal is consistent with data processing artifacts that masquerade as an oscillating signal in a truly $\Lambda$CDM cosmology.  
 
 \section{Acknowledgments} We found the work of \cite{VanderPlas2018} particularly helpful in understanding the importance of being cautious about the potential impact of processing artifacts. SB and CS are supported by Harvard University and the US Department of Energy under grant DE-SC0007881.  DS is supported by DOE grant DE-SC0010007 and the David and Lucile Packard Foundation. DS is supported in part by NASA under Contract No. NNG17PX03C issued through the WFIRST Science Investigation Teams Programme.

\newpage
\bibliography{myRingermacher.bib}

\begin{thebibliography}{24}
\providecommand{\natexlab}[1]{#1}
\providecommand{\url}[1]{\texttt{#1}}
\expandafter\ifx\csname urlstyle\endcsname\relax
  \providecommand{\doi}[1]{doi: #1}\else
  \providecommand{\doi}{doi: \begingroup \urlstyle{rm}\Url}\fi

\bibitem[{Barenboim} et~al.(2005){Barenboim}, {Requejo}, and
  {Quigg}]{Barenboim2005}
G.~{Barenboim}, O.~M. {Requejo}, and C.~{Quigg}.
\newblock {Undulant Universe: Expansion with alternating eras of acceleration
  and deceleration}.
\newblock \emph{\prd}, 71\penalty0 (6):\penalty0 063533, Mar. 2005.
\newblock \doi{10.1103/PhysRevD.71.063533}.

\bibitem[Brownsberger et~al.(2019)Brownsberger, Stubbs, and
  Scolnic]{Brownsberger2019}
S.~R. Brownsberger, C.~W. Stubbs, and D.~M. Scolnic.
\newblock Constraining temporal oscillations of cosmological parameters using
  {SNe} ia.
\newblock \emph{The Astrophysical Journal}, 875\penalty0 (1):\penalty0 34, apr
  2019.
\newblock \doi{10.3847/1538-4357/ab0c09}.
\newblock URL \url{https://doi.org/10.3847%2F1538-4357%2Fab0c09}.

\bibitem[{Conley} et~al.(2011){Conley}, {Guy}, {Sullivan}, {Regnault},
  {Astier}, {Balland}, {Basa}, {Carlberg}, {Fouchez}, {Hardin}, {Hook},
  {Howell}, {Pain}, {Palanque-Delabrouille}, {Perrett}, {Pritchet}, {Rich},
  {Ruhlmann-Kleider}, {Balam}, {Baumont}, {Ellis}, {Fabbro}, {Fakhouri},
  {Fourmanoit}, {Gonz{\'a}lez-Gait{\'a}n}, {Graham}, {Hudson}, {Hsiao},
  {Kronborg}, {Lidman}, {Mourao}, {Neill}, {Perlmutter}, {Ripoche}, {Suzuki},
  and {Walker}]{Conley2011}
A.~{Conley}, J.~{Guy}, M.~{Sullivan}, N.~{Regnault}, P.~{Astier}, C.~{Balland},
  S.~{Basa}, R.~G. {Carlberg}, D.~{Fouchez}, D.~{Hardin}, I.~M. {Hook}, D.~A.
  {Howell}, R.~{Pain}, N.~{Palanque-Delabrouille}, K.~M. {Perrett}, C.~J.
  {Pritchet}, J.~{Rich}, V.~{Ruhlmann-Kleider}, D.~{Balam}, S.~{Baumont}, R.~S.
  {Ellis}, S.~{Fabbro}, H.~K. {Fakhouri}, N.~{Fourmanoit},
  S.~{Gonz{\'a}lez-Gait{\'a}n}, M.~L. {Graham}, M.~J. {Hudson}, E.~{Hsiao},
  T.~{Kronborg}, C.~{Lidman}, A.~M. {Mourao}, J.~D. {Neill}, S.~{Perlmutter},
  P.~{Ripoche}, N.~{Suzuki}, and E.~S. {Walker}.
\newblock {Supernova Constraints and Systematic Uncertainties from the First
  Three Years of the Supernova Legacy Survey}.
\newblock \emph{\apjs}, 192\penalty0 (1):\penalty0 1, Jan. 2011.
\newblock \doi{10.1088/0067-0049/192/1/1}.

\bibitem[{Daly} and {Djorgovski}(2004)]{Daly2004}
R.~A. {Daly} and S.~G. {Djorgovski}.
\newblock {Direct Determination of the Kinematics of the Universe and
  Properties of the Dark Energy as Functions of Redshift}.
\newblock \emph{\apj}, 612\penalty0 (2):\penalty0 652--659, Sept. 2004.
\newblock \doi{10.1086/422673}.

\bibitem[Feng and Li(2006)]{Feng2006}
B.~Feng and M.~Li.
\newblock Oscillating quintom and the recurrent universe.
\newblock \emph{Physics Letters B}, 634\penalty0 (2):\penalty0 101 -- 105,
  2006.
\newblock ISSN 0370-2693.
\newblock \doi{https://doi.org/10.1016/j.physletb.2006.01.066}.
\newblock URL
  \url{http://www.sciencedirect.com/science/article/pii/S0370269306001407}.

\bibitem[{Hicken} et~al.(2009){Hicken}, {Wood-Vasey}, {Blondin}, {Challis},
  {Jha}, {Kelly}, {Rest}, and {Kirshner}]{Hicken2009}
M.~{Hicken}, W.~M. {Wood-Vasey}, S.~{Blondin}, P.~{Challis}, S.~{Jha}, P.~L.
  {Kelly}, A.~{Rest}, and R.~P. {Kirshner}.
\newblock {Improved Dark Energy Constraints from
  \raisebox{-0.5ex}\textasciitilde100 New CfA Supernova Type Ia Light Curves}.
\newblock \emph{\apj}, 700\penalty0 (2):\penalty0 1097--1140, Aug. 2009.
\newblock \doi{10.1088/0004-637X/700/2/1097}.

\bibitem[Jain et~al.(2007)Jain, Dev, Alcaniz, and et~al.]{Jain2007}
D.~Jain, A.~Dev, J.~Alcaniz, and et~al.
\newblock Cosmological bounds on oscillating dark energy models.
\newblock \emph{Physics Letters B}, 656\penalty0 (1):\penalty0 15 -- 18, 2007.
\newblock ISSN 0370-2693.
\newblock \doi{https://doi.org/10.1016/j.physletb.2007.09.023}.
\newblock URL
  \url{http://www.sciencedirect.com/science/article/pii/S0370269307011598}.

\bibitem[{Kowalski} et~al.(2008){Kowalski}, {Rubin}, {Aldering}, {Agostinho},
  {Amadon}, {Amanullah}, {Balland }, {Barbary}, {Blanc}, {Challis}, {Conley},
  {Connolly}, {Covarrubias}, {Dawson}, {Deustua}, {Ellis}, {Fabbro}, {Fadeyev},
  {Fan}, {Farris}, {Folatelli}, {Frye}, {Garavini}, {Gates}, {Germany},
  {Goldhaber}, {Goldman}, {Goobar}, {Groom}, {Haissinski}, {Hardin}, {Hook},
  {Kent}, {Kim}, {Knop}, {Lidman}, {Linder}, {Mendez}, {Meyers}, {Miller},
  {Moniez}, {Mour{\~a}o}, {Newberg}, {Nobili}, {Nugent}, {Pain}, {Perdereau},
  {Perlmutter}, {Phillips}, {Prasad}, {Quimby}, {Regnault}, {Rich},
  {Rubenstein}, {Ruiz-Lapuente}, {Santos}, {Schaefer}, {Schommer}, {Smith},
  {Soderberg}, {Spadafora}, {Strolger}, {Strovink}, {Suntzeff}, {Suzuki},
  {Thomas}, {Walton}, {Wang}, {Wood-Vasey}, and {Yun}]{Kowalski2008}
M.~{Kowalski}, D.~{Rubin}, G.~{Aldering}, R.~J. {Agostinho}, A.~{Amadon},
  R.~{Amanullah}, C.~{Balland }, K.~{Barbary}, G.~{Blanc}, P.~J. {Challis},
  A.~{Conley}, N.~V. {Connolly}, R.~{Covarrubias}, K.~S. {Dawson}, S.~E.
  {Deustua}, R.~{Ellis}, S.~{Fabbro}, V.~{Fadeyev}, X.~{Fan}, B.~{Farris},
  G.~{Folatelli}, B.~L. {Frye}, G.~{Garavini}, E.~L. {Gates}, L.~{Germany},
  G.~{Goldhaber}, B.~{Goldman}, A.~{Goobar}, D.~E. {Groom}, J.~{Haissinski},
  D.~{Hardin}, I.~{Hook}, S.~{Kent}, A.~G. {Kim}, R.~A. {Knop}, C.~{Lidman},
  E.~V. {Linder}, J.~{Mendez}, J.~{Meyers}, G.~J. {Miller}, M.~{Moniez}, A.~M.
  {Mour{\~a}o}, H.~{Newberg}, S.~{Nobili}, P.~E. {Nugent}, R.~{Pain},
  O.~{Perdereau}, S.~{Perlmutter}, M.~M. {Phillips}, V.~{Prasad}, R.~{Quimby},
  N.~{Regnault}, J.~{Rich}, E.~P. {Rubenstein}, P.~{Ruiz-Lapuente}, F.~D.
  {Santos}, B.~E. {Schaefer}, R.~A. {Schommer}, R.~C. {Smith}, A.~M.
  {Soderberg}, A.~L. {Spadafora}, L.~G. {Strolger}, M.~{Strovink}, N.~B.
  {Suntzeff}, N.~{Suzuki}, R.~C. {Thomas}, N.~A. {Walton}, L.~{Wang}, W.~M.
  {Wood-Vasey}, and J.~L. {Yun}.
\newblock {Improved Cosmological Constraints from New, Old, and Combined
  Supernova Data Sets}.
\newblock \emph{\apj}, 686\penalty0 (2):\penalty0 749--778, Oct. 2008.
\newblock \doi{10.1086/589937}.

\bibitem[Lazkoz et~al.(2010)Lazkoz, Salzano, Sendra, and et~al.]{Lazkoz2010}
R.~Lazkoz, V.~Salzano, I.~Sendra, and et~al.
\newblock Oscillations in the dark energy equation of state: New mcmc lessons.
\newblock \emph{Physics Letters B}, 694\penalty0 (3):\penalty0 198 -- 208,
  2010.
\newblock ISSN 0370-2693.
\newblock \doi{https://doi.org/10.1016/j.physletb.2010.10.002}.
\newblock URL
  \url{http://www.sciencedirect.com/science/article/pii/S0370269310011871}.

\bibitem[Liu and Li(2009)]{Liu2009}
J.~Liu and H.~Li.
\newblock Testing oscillating primordial spectrum and oscillating dark energy
  with astronomical observations.
\newblock \emph{Journal of Cosmology and Astroparticle Physics}, 2009\penalty0
  (07):\penalty0 017, 2009.
\newblock URL \url{http://stacks.iop.org/1475-7516/2009/i=07/a=017}.

\bibitem[{Mortonson} et~al.(2013){Mortonson}, {Weinberg}, and
  {White}]{Mortonson2013}
M.~J. {Mortonson}, D.~H. {Weinberg}, and M.~{White}.
\newblock {Dark Energy: A Short Review}.
\newblock \emph{arXiv e-prints}, art. arXiv:1401.0046, Dec. 2013.

\bibitem[{Perlmutter} et~al.(1999){Perlmutter}, {Aldering}, {Goldhaber},
  {Knop}, {Nugent}, {Castro}, {Deustua}, {Fabbro}, {Goobar}, {Groom}, {Hook},
  {Kim}, {Kim}, {Lee}, {Nunes}, {Pain}, {Pennypacker}, {Quimby}, {Lidman},
  {Ellis}, {Irwin}, {McMahon}, {Ruiz-Lapuente}, {Walton}, {Schaefer}, {Boyle},
  {Filippenko}, {Matheson}, {Fruchter}, {Panagia}, {Newberg}, {Couch}, and
  {Project}]{Perlmutter1999}
S.~{Perlmutter}, G.~{Aldering}, G.~{Goldhaber}, R.~A. {Knop}, P.~{Nugent},
  P.~G. {Castro}, S.~{Deustua}, S.~{Fabbro}, A.~{Goobar}, D.~E. {Groom}, I.~M.
  {Hook}, A.~G. {Kim}, M.~Y. {Kim}, J.~C. {Lee}, N.~J. {Nunes}, R.~{Pain},
  C.~R. {Pennypacker}, R.~{Quimby}, C.~{Lidman}, R.~S. {Ellis}, M.~{Irwin},
  R.~G. {McMahon}, P.~{Ruiz-Lapuente}, N.~{Walton}, B.~{Schaefer}, B.~J.
  {Boyle}, A.~V. {Filippenko}, T.~{Matheson}, A.~S. {Fruchter}, N.~{Panagia},
  H.~J.~M. {Newberg}, W.~J. {Couch}, and T.~S.~C. {Project}.
\newblock {Measurements of {\ensuremath{\Omega}} and {\ensuremath{\Lambda}}
  from 42 High-Redshift Supernovae}.
\newblock \emph{\apj}, 517\penalty0 (2):\penalty0 565--586, June 1999.
\newblock \doi{10.1086/307221}.

\bibitem[Peter(2012)]{Peter2012}
A.~H.~G. Peter.
\newblock Dark matter: A brief review.
\newblock \emph{arXiv}, 2012.

\bibitem[Riess et~al.(1998)Riess, Filippenko, Challis, Clocchiatti, Diercks,
  Garnavich, Gilliland, Hogan, Jha, Kirshner, Leibundgut, Phillips, Reiss,
  Schmidt, Schommer, Smith, Spyromilio, Stubbs, Suntzeff, and Tonry]{Riess1998}
A.~G. Riess, A.~V. Filippenko, P.~Challis, A.~Clocchiatti, A.~Diercks, P.~M.
  Garnavich, R.~L. Gilliland, C.~J. Hogan, S.~Jha, R.~P. Kirshner,
  B.~Leibundgut, M.~M. Phillips, D.~Reiss, B.~P. Schmidt, R.~A. Schommer, R.~C.
  Smith, J.~Spyromilio, C.~Stubbs, N.~B. Suntzeff, and J.~Tonry.
\newblock Observational evidence from supernovae for an accelerating universe
  and a cosmological constant.
\newblock \emph{The Astronomical Journal}, 116\penalty0 (3):\penalty0
  1009--1038, sep 1998.
\newblock \doi{10.1086/300499}.
\newblock URL \url{https://doi.org/10.1086%2F300499}.

\bibitem[{Riess} et~al.(2004){Riess}, {Strolger}, {Tonry}, {Casertano},
  {Ferguson}, {Mobasher}, {Challis}, {Filippenko}, {Jha}, {Li}, {Chornock},
  {Kirshner}, {Leibundgut}, {Dickinson}, {Livio}, {Giavalisco}, {Steidel},
  {Ben{\'\i}tez}, and {Tsvetanov}]{Reiss2004}
A.~G. {Riess}, L.-G. {Strolger}, J.~{Tonry}, S.~{Casertano}, H.~C. {Ferguson},
  B.~{Mobasher}, P.~{Challis}, A.~V. {Filippenko}, S.~{Jha}, W.~{Li},
  R.~{Chornock}, R.~P. {Kirshner}, B.~{Leibundgut}, M.~{Dickinson}, M.~{Livio},
  M.~{Giavalisco}, C.~C. {Steidel}, T.~{Ben{\'\i}tez}, and Z.~{Tsvetanov}.
\newblock {Type Ia Supernova Discoveries at z \&gt; 1 from the Hubble Space
  Telescope: Evidence for Past Deceleration and Constraints on Dark Energy
  Evolution}.
\newblock \emph{\apj}, 607\penalty0 (2):\penalty0 665--687, June 2004.
\newblock \doi{10.1086/383612}.

\bibitem[{Riess} et~al.(2007){Riess}, {Strolger}, {Casertano}, {Ferguson},
  {Mobasher}, {Gold}, {Challis}, {Filippenko}, {Jha}, {Li}, {Tonry}, {Foley},
  {Kirshner}, {Dickinson}, {MacDonald}, {Eisenstein}, {Livio}, {Younger}, {Xu},
  {Dahl{\'e}n}, and {Stern}]{Reiss2007}
A.~G. {Riess}, L.-G. {Strolger}, S.~{Casertano}, H.~C. {Ferguson},
  B.~{Mobasher}, B.~{Gold}, P.~J. {Challis}, A.~V. {Filippenko}, S.~{Jha},
  W.~{Li}, J.~{Tonry}, R.~{Foley}, R.~P. {Kirshner}, M.~{Dickinson},
  E.~{MacDonald}, D.~{Eisenstein}, M.~{Livio}, J.~{Younger}, C.~{Xu},
  T.~{Dahl{\'e}n}, and D.~{Stern}.
\newblock {New Hubble Space Telescope Discoveries of Type Ia Supernovae at z
  \&gt;= 1: Narrowing Constraints on the Early Behavior of Dark Energy}.
\newblock \emph{\apj}, 659\penalty0 (1):\penalty0 98--121, Apr. 2007.
\newblock \doi{10.1086/510378}.

\bibitem[Ringermacher and Mead(2015)]{Ringermacher2015}
H.~I. Ringermacher and L.~R. Mead.
\newblock {OBSERVATION} {OF} {DISCRETE} {OSCILLATIONS} {IN} a
  {MODEL}-{INDEPENDENT} {PLOT} {OF} {COSMOLOGICAL} {SCALE} {FACTOR} {VERSUS}
  {LOOKBACK} {TIME} {AND} {SCALAR} {FIELD} {MODEL}.
\newblock \emph{The Astronomical Journal}, 149\penalty0 (4):\penalty0 137, mar
  2015.
\newblock \doi{10.1088/0004-6256/149/4/137}.
\newblock URL \url{https://doi.org/10.1088%2F0004-6256%2F149%2F4%2F137}.

\bibitem[Ringermacher and Mead(2020)]{Ringermacher2020}
H.~I. Ringermacher and L.~R. Mead.
\newblock {Reaffirmation of cosmological oscillations in the scale factor from
  the Pantheon compilation of 1048 Type Ia supernovae}.
\newblock \emph{Monthly Notices of the Royal Astronomical Society},
  494\penalty0 (2):\penalty0 2158--2165, 04 2020.
\newblock ISSN 0035-8711.
\newblock \doi{10.1093/mnras/staa872}.
\newblock URL \url{https://doi.org/10.1093/mnras/staa872}.

\bibitem[Scolnic et~al.(2018)Scolnic, Jones, Rest, Pan, Chornock, Foley, Huber,
  Kessler, Narayan, Riess, Rodney, Berger, Brout, Challis, Drout, Finkbeiner,
  Lunnan, Kirshner, Sanders, Schlafly, Smartt, Stubbs, Tonry, Wood-Vasey,
  Foley, Hand, Johnson, Burgett, Chambers, Draper, Hodapp, Kaiser, Kudritzki,
  Magnier, Metcalfe, Bresolin, Gall, Kotak, McCrum, and Smith]{Scolnic2018}
D.~M. Scolnic, D.~O. Jones, A.~Rest, Y.~C. Pan, R.~Chornock, R.~J. Foley, M.~E.
  Huber, R.~Kessler, G.~Narayan, A.~G. Riess, S.~Rodney, E.~Berger, D.~J.
  Brout, P.~J. Challis, M.~Drout, D.~Finkbeiner, R.~Lunnan, R.~P. Kirshner,
  N.~E. Sanders, E.~Schlafly, S.~Smartt, C.~W. Stubbs, J.~Tonry, W.~M.
  Wood-Vasey, M.~Foley, J.~Hand, E.~Johnson, W.~S. Burgett, K.~C. Chambers,
  P.~W. Draper, K.~W. Hodapp, N.~Kaiser, R.~P. Kudritzki, E.~A. Magnier,
  N.~Metcalfe, F.~Bresolin, E.~Gall, R.~Kotak, M.~McCrum, and K.~W. Smith.
\newblock The complete light-curve sample of spectroscopically confirmed {SNe}
  ia from pan-{STARRS}1 and cosmological constraints from the combined pantheon
  sample.
\newblock \emph{The Astrophysical Journal}, 859\penalty0 (2):\penalty0 101, may
  2018.
\newblock URL \url{https://doi.org/10.3847%2F1538-4357%2Faab9bb}.

\bibitem[{VanderPlas}(2018)]{VanderPlas2018}
J.~T. {VanderPlas}.
\newblock {Understanding the Lomb-Scargle Periodogram}.
\newblock \emph{\apjs}, 236\penalty0 (1):\penalty0 16, May 2018.
\newblock \doi{10.3847/1538-4365/aab766}.

\bibitem[{Verde} et~al.(2019){Verde}, {Treu}, and {Riess}]{Verde2019}
L.~{Verde}, T.~{Treu}, and A.~G. {Riess}.
\newblock {Tensions between the early and late Universe}.
\newblock \emph{Nature Astronomy}, 3:\penalty0 891--895, Sept. 2019.
\newblock \doi{10.1038/s41550-019-0902-0}.

\bibitem[{Wang} et~al.(2017){Wang}, {Zhu}, and {Unruh}]{Wang2017}
Q.~{Wang}, Z.~{Zhu}, and W.~G. {Unruh}.
\newblock {How the huge energy of quantum vacuum gravitates to drive the slow
  accelerating expansion of the Universe}.
\newblock \emph{\prd}, 95\penalty0 (10):\penalty0 103504, May 2017.
\newblock \doi{10.1103/PhysRevD.95.103504}.

\bibitem[Weinberg et~al.(2015)Weinberg, Bullock, Governato, de~Naray, and
  Peter]{Weinberg2015}
D.~H. Weinberg, J.~S. Bullock, F.~Governato, R.~K. de~Naray, and A.~H.~G.
  Peter.
\newblock Cold dark matter: Controversies on small scales.
\newblock \emph{PNAS}, 122\penalty0 (40), 2015.

\bibitem[Xia et~al.(2005)Xia, Feng, Zhang, and et~al.]{Xia2005}
J.~Xia, B.~Feng, X.~Zhang, and et~al.
\newblock Constraints on oscillating quintom from supernova, microwave
  background and galaxy clustering.
\newblock \emph{Modern Physics Letters A}, 20\penalty0 (31):\penalty0
  2409--2416, 2005.
\newblock \doi{10.1142/S0217732305017445}.
\newblock URL
  \url{https://www.worldscientific.com/doi/abs/10.1142/S0217732305017445}.

\end{thebibliography}
\bibliographystyle{abbrvnat}

\end{document}